\documentstyle[aps,prl,multicol]{revtex} 

\newcommand{\e}{\epsilon}

\newcommand{\rem}[1]{}

\begin{document}
\title{An Integrable Shallow Water Equation
with Linear and Nonlinear Dispersion}

\author{\vspace{-9mm}
\begin{multicols}{3}
Holger R. Dullin
   \\ Mathematical Sciences
   \\ Loughborough University
   \\ Loughborough, UK
   \\{\small h.r.dullin@lboro.ac.uk}
   \and\\  \vspace{1cm}
Georg Gottwald
   \\Mathematics and Statistics
   \\University of Surrey
   \\Guildford, Surrey
   \\{\small g.gottwald@eim.surrey.ac.uk}
\and\\  \vspace{1cm}
Darryl D. Holm
\\ CNLS and Theoretical Division
\\ Los Alamos National Laboratory
\\ Los Alamos, NM 87545\\
{\small dholm@lanl.gov}
\end{multicols}
}

\maketitle

\vspace{-5mm}
\begin{abstract}

We study a class of 1+1 quadratically nonlinear water wave equations
that combines the linear dispersion of the Korteweg-deVries (KdV)
equation with the nonlinear/nonlocal dispersion of the Camassa-Holm
(CH) equation, yet still preserves integrability via the inverse
scattering transform (IST) method.
This IST-integrable class of equations contains both the KdV equation
and the CH equation as limiting cases. It arises as the compatibility
condition for a second order isospectral eigenvalue problem and a
first order equation for the evolution of its eigenfunctions.
This integrable equation is shown to be a shallow water wave equation
derived by asymptotic expansion at one order higher approximation
than KdV. We compare its traveling wave solutions to KdV solitons.



\end{abstract}

\vskip 4mm

\begin{multicols}{2}


Water wave theory first introduced solitons as solutions of
unidirectional nonlinear wave equations, obtained via asymptotic
expansions around simple wave motion of the Euler equations for
shallow water  in a particular Galilean frame
\cite{Whitham[1974]}. Later developments identified some of these water
wave equations as completely integrable Hamiltonian systems solvable by the
inverse
scattering transform (IST) method, see, e.g., \cite{Ablowitz&Segur[1981]}.
We shall discuss the following 1+1 quadratically nonlinear equation in
this class for unidirectional water waves with fluid velocity $u(x,t)$,
\begin{equation}\label{CH-BV}
m_t + c_0 u_x + u\,m_x + 2m\,u_x = -\,\gamma\, u_{xxx}
\,.
\end{equation}
Here $m=u-\alpha^{\,2}u_{xx}$ is a momentum variable, partial
derivatives are denoted by subscripts, the constants $\alpha^{\,2}$ and
$\gamma/c_0$ are squares of length scales, and $c_0=\sqrt{gh}$ is the linear
wave speed for undisturbed water at rest at spatial infinity, where
$u$ and $m$ are taken to vanish. (Any constant value $u=u_0$ is also a
solution.) Equation (\ref{CH-BV}) was first derived by using asymptotic
expansions directly in the Hamiltonian for Euler's equations in the
shallow water regime and was thereby shown to be bi-Hamiltonian and, thus,
IST-integrable in \cite{CH[1993]}. Before \cite{CH[1993]}, classes of
integrable equations similar to (\ref{CH-BV}) were known to be derivable
from the theory of hereditary symmetries, \cite{Fokas&Fuchssteiner}.
However, these were not derived physically as water wave equations and
their solution properties were not studied before \cite{CH[1993]}. See
\cite{Fuchssteiner} for an insightful discussion of how the integrable
equation (\ref{CH-BV}) relates to the theory of hereditary symmetries.

The interplay between the local and nonlocal linear dispersion in this
equation is evident in its phase velocity relation,
${\omega}/{k}
=
(c_0 - {\gamma}\, k^2)\,/(1 + \alpha^{\,2} k^2)$,
for waves with frequency $\omega$ and wave number $k$ linearized
around ${u}=0$. At low wave numbers, the
constant dispersion parameters
$\alpha^{\,2}$ and $\gamma$ perform rather similar functions.
At high wave numbers, however, the parameter $\alpha^{\,2}$ properly
keeps the phase speed of the wave from becoming unbounded. The phase
speed lies in the band $\omega/k\in(-\,\gamma/\alpha^{\,2}, c_0)$.
Longer linear waves are the faster provided
$\gamma + c_0 \alpha^2 \ge 0$.

Equation (\ref{CH-BV}) is not Galilean invariant. Upon shifting the
velocity variable by $u_0$ and moving into a Galilean frame $\xi=x-ct$
with velocity $c$, so that
$u(x,t)=\tilde{u}(\xi,t)+c+u_0$, this equation transforms to
\begin{equation}\label{CH-BV-xform}
\tilde{m}_t + \tilde{u}\,\tilde{m}_\xi + 2\tilde{m}\,\tilde{u}_\xi
=
-\tilde c_0\, u_\xi - \tilde{\gamma}\, u_{\xi\xi\xi}
\,,
\nonumber
\end{equation}
with $\tilde{c_0}=(c_0 + 2c + 3u_0)$,
$\tilde{\gamma}=(\gamma - u_0\alpha^{\,2})$ and appropriately altered
boundary conditions  at spatial infinity.
Hence, we must regard equation (\ref{CH-BV}) as a {\it family} of
equations whose linear dispersion parameters $c_0$, $\gamma$ depend on the
appropriate choice of Galilean frame and boundary conditions. The
parameters $c_0$ and $\gamma$ may even be {\it removed} by making
such a choice \cite{CH[1993]}. In the following we will only use
transformations that leave the boundary condition $u=0$ at spatial
infinity invariant.

This paper reports two main results. First, we identify how the dispersion
coefficients for the linearized water waves appear as {\it parameters} in
the isospectral problem for this IST-integrable equation. We determine
how the linear dispersion parameters $\alpha$, $c_0$ and $\gamma$ in
(\ref{CH-BV}) affect the isospectral content of its soliton solutions and
the shape of its traveling waves. Second, we rederive equation
(\ref{CH-BV}) by using a certain {\it nonlocal} form of shallow water wave
asymptotics that is correct to one order higher than for KdV.
This new derivation and the analysis we present here attaches additional
physical meaning to equation (\ref{CH-BV}) in the context of asymptotics
for shallow water wave equations. In fact, it even allows us to optimize
its dispersion coefficients for maximal accuracy.

Equation (\ref{CH-BV}) restricts to two separately
integrable soliton equations for water waves. When $\alpha^{\,2}\to0$
this equation becomes the Korteweg-deVries (KdV) equation,
\begin{equation}\label{KdV-eqn}
u_t + c_0 u_x + 3u\,u_x = -\,\gamma\, u_{xxx}
\,,
\nonumber
\end{equation}
which for $c_0=0$ has the famous soliton solution
$u(x,t)=u_0\,{\rm sech}^2((x-ct)\sqrt{u_0/\gamma}/2)$, $c = c_0+u_0$
see, e.g., \cite{Ablowitz&Segur[1981]}.
Instead, taking $\gamma\to0$ in
equation (\ref{CH-BV}) implies the Camassa-Holm (CH) equation,
%
\[
u_t + c_0 u_x - \alpha^{\,2}\,u_{xxt} + 3u\,u_x
= \alpha^{\,2}\Big(2u_xu_{xx} + uu_{xxx}\Big)
\,,
\]
%
which for $c_0=0$ has ``peakon'' soliton solutions
$u(x,t)=ce^{-|x-ct|}$ discovered and analyzed in \cite{CH[1993]}.

Equation (\ref{CH-BV}) also arises as the Euler-Poincar\'e
equation  for an averaged Lagrangian, \cite{EP-footnote}, \cite{HMR-98ab}.
Its solutions describe geodesic motion with respect to the $H_1$ metric of
$u$ on the Bott-Virasoro Lie group \cite{Misiolek[1998]}. The KdV equation
arises the same way for the $L_2$ metric, see, e.g., \cite{EP-footnote}.

Equation (\ref{CH-BV}) is nonlocal, as is seen by
rewriting it as
%
\[
(1-\alpha^{\,2}\partial_x^2)(u_t + u\,u_x)
=-
\partial_x
\!
\left(
\!\!
\left(u+\frac{c_0}{2}\right)^2
+
\frac{\alpha^{\,2} u_x^2 }{2}
+
\gamma{u}_{xx}
\!\!
\right)
\]
%
%
and inverting the Helmholtz operator using its Green's function
relation $(1-\alpha^{\,2}\partial_x^2)\,e^{-\,|\,x\,|\,/\alpha\,}
=2\delta(x/\alpha)$. Dropping nonlinear terms on the right side and
setting $\gamma=0$ recovers the equation studied by Fornberg and Whitham
\cite{Fornberg-Whitham[1978]}.
Thus, equation (\ref{CH-BV}) contains three simpler water wave
equations, two of which are known to be integrable.


{\it Equation (\ref{CH-BV}) is bi-Hamiltonian and hence, isospectral.}
The term bi-Hamiltonian means the equation may be written in two
compatible Hamiltonian ways, namely as
$m_t
=
-B_2 ({\delta H_1}/{\delta m})
=
-B_1 ({\delta H_2}/{\delta m})
$
with
\begin{eqnarray}\label{CH-BV-bi-Ham}
H_1
&=&
\frac{1}{2}\int u^{\,2} + \alpha^{\,2} u_x^{\,2}\
dx
\,,\quad
B_2
=
\partial_xm + m\partial_x + \gamma\,\partial_x^3
\nonumber\\
H_2
&=&
\frac{1}{2}\int
u^{\,3}
+
\alpha^{\,2} u\,u_x^{\,2}
-
\gamma\, u_x^{\,2}\
dx
\,,\quad
B_1
=
\partial_x - \alpha^{\,2} \partial_x^3
\,.\nonumber
\end{eqnarray}
These bi-Hamiltonian restrict properly to those for KdV when
$\alpha^{\,2}\to0$, and to those for CH when $\gamma\to0$. Compatibility
of $B_1$ and $B_2$ is assured, since $\partial_xm + m\partial_x$,
$\partial_x$ and $\partial_x^3$ are all mutually compatible
Hamiltonian operators, see, e.g., \cite{Olver[1993]}. From this
viewpoint, $\gamma$ and $\alpha^2$ are deformation parameters
for the {\it Riemann equation}, $u_t+3uu_x=0$.
No further deformations of these Hamiltonian operators involving higher
order partial derivatives would be compatible with $B_2$
\cite{Olver[2000]}. A related approach is given in
\cite{Fokas[1995],Fokas-Olver-Rosenau[1996]}.

By the standard Gelfand-Dorfman theory \cite{SquaredEV-foot}, its
bi-Hamiltonian property implies that the nonlinear equation
(\ref{CH-BV}) arises as a compatibility condition for two linear
equations, namely, the {\it isospectral eigenvalue problem},
\begin{equation}\label{CH-BV-isospectral}\hspace{-1cm}
\lambda\, \Big(\frac{1}{4} - \alpha^{\,2}\partial_x^{\,2}\Big)\psi
=
  \left( \frac{c_0}{4} + \frac{m(x,t)}{2}
+
\gamma\,\partial_x^{\,2}\right)\psi
\,,
\end{equation}
and the {\it evolution equation} for the eigenfunction $\psi$,
\begin{equation}\label{CH-BV-evol}
\psi_t
=
-(u+\lambda)\,\psi_x + \frac{1}{2}u_x\,\psi
\,.
\nonumber
\end{equation}
Compatibility of these linear equations $(\psi_{xxt}=\psi_{txx})$ and
isospectrality $(d\lambda/dt=0)$ imply equation (\ref{CH-BV}).
Consequently, the nonlinear water wave equation (\ref{CH-BV}) admits
the IST method for the solution of its initial value problem, just as
the KdV and CH  equations do. In fact, the isospectral problem for
equation (\ref{CH-BV}) restricts to the isospectral problem for KdV
(i.e., the Schr\"odinger equation) when $\alpha^{\,2}\to0$ and it
restricts to the isospectral problem for CH discovered in
\cite{CH[1993]} when $\gamma\to0$.

The IST method for solving the KdV equation was introduced in
\cite{GGKM[1967]}. The isospectral problem associated with solving the CH
equation by IST is treated in \cite{Fokas[1995],Bealsetal[1998]}. The
combined eigenvalue problem (\ref{CH-BV-isospectral}) parameterized by
$\alpha$, $c_0$ and $\gamma$ is a new variant that retains many of the same
properties as occur for restrictions to the zero values of these
parameters. Thus, the IST approach based on this isospectral eigenvalue
problem will illuminate the dependence of the solution behavior of equation
(\ref{CH-BV}) on its dispersion parameters $\alpha$, $c_0$ and
$\gamma$, in comparison with the solution behavior of its two limiting
integrable water wave equations, KdV and CH.

Defining a new spectral parameter
$\chi^{-2} = \gamma + \lambda \alpha^2$ yields
a spectral problem in the same form as for CH,
\[
      \left( \partial_x^2 - \frac{1}{4\alpha^2} + \chi^2
          \left(\frac{m}{2} + \frac{\gamma + c_0
\alpha^2}{4\alpha^2}\right)\right)\psi = 0
\,,\]
which is analyzed and discussed in \cite{Fokas[1995],Bealsetal[1998]}.
The continuous spectrum lies in $\chi^{-2} \in[0, \gamma + c_0 \alpha^2)$.
In this form of the isospectral equation, the limit
to KdV is singular.


{\it Spectral content and solution behavior.} Provided $m$
decreases sufficiently rapidly at spatial infinity, equation
(\ref{CH-BV-isospectral}) has both continuous and discrete spectra.
These spectral components correspond to the two different types of
solution behavior available for equation (\ref{CH-BV}). The continuous
spectrum of the isospectral eigenvalue problem (\ref{CH-BV-isospectral})
spans the band of allowed linearized phase speeds, namely
$\lambda\in(-\,\gamma/\alpha^{\,2},c_0)$.
This continuous spectrum corresponds to radiation (linear waves). The
discrete spectrum of (\ref{CH-BV-isospectral}) lies above this band,
with $\lambda>c_0\ge0$. The discrete spectrum corresponds to
the soliton sector of the solution space. This is also what is seen in
numerical computations \cite{DGH[2001]}. In the limit that both $c_0\to0$
and $\gamma\to0$, the corresponding isospectral problem for the CH equation
has purely discrete spectrum representing only peakon solutions
\cite{CH[1993]}.

{\it The derivation from shallow water wave asymptotics} is similar to
that of the KdV equation, except that we keep terms of higher order
in the small parameters $\e_1 = a/h$ and $\e_2 = h^2/l^2$ where
$\e_1 > \e_2 > \e_1^2$ and $a$, $h$, and $l$ are the wave
amplitude, the average total water depth, and a typical horizontal
length scale, respectively. We shall sketch the derivation here and give
details elsewhere \cite{DGH[2001]}. We start with the equations for an
inviscid, incompressible, and irrotational fluid moving under gravity with
an upper free surface, as, e.g., in \cite{Whitham[1974]}.
The velocity potential is expanded at a fixed height $z_0$
\cite{Nwogu[1993]}. 
This is preferable to the usual depth averaging because
it allows a parameter freedom in the expansion that may later be
optimized. The intermediate steps of this calculation all involve the
chosen height $z_0$. However, the equation for the elevation
of the surface $\eta$ is the same as in the standard approach (see,
e.g., \cite{Whitham[1974]}).
Length is measured in terms of $l$, height in $h$ and time in $l/c_0$. The
elevation $\eta$ is scaled with $a$ and fluid velocity $u$ is scaled with
$c_0 a / h$. In nondimensional variables one finds
\begin{eqnarray}
\label{asy1}
0 &=& \eta_t+\eta_x+\frac{3}{2}\e_1\eta \eta_x
-\frac{3}{8}\epsilon_1^2\eta^2\eta_x+\frac{1}{6}\epsilon_2\eta_{xxx}+\\
    && \quad
+\epsilon_1\epsilon_2\left(\frac{23}{24}\eta_x\eta_{xx}
+\frac{5}{12}\eta\eta_{xxx}\right)\;.
\nonumber
\end{eqnarray}
Transforming (\ref{asy1}) to the velocity $u$ gives  equation
(\ref{CH-BV}), when we expand $\eta$ in a series in terms of the fluid
velocity $u = \phi_x(x,t;z_0)$ as
\begin{equation} \label{Kodama}
   \eta = u + \e_1\left( \alpha_1 u^2 + \alpha_2 u_x \partial_x^{-1}
u\right) + \e_2 \beta u_{xx} \,.
\end{equation}
In this transformation we include a crucial nonlocal term that was first
introduced by Kodama in \cite{Kodama[1985]}.
Usually the coefficients of this expansion are constrained by the
requirement to match the velocity $u=\phi_x$ averaged over the fluid
column.  (The derivation in \cite{Fokas[1995]}
apparently ignores this constraint.)
Considering the fluid velocity at height $z_0$ removes this constraint.
We introduce another optimization parameter by splitting the higher
order time derivative $\e_2 \beta u_{xxt}$ obtained by substituting
(\ref{Kodama}) into (\ref{asy1})
\[
\beta u_{xxt} = -\nu u_{xxt} - (\beta + \nu)\left( u_{xxx} +
\frac{3\e_1}{2}(3u_xu_{xx} + u u_{xxx}) \right)
\]
which is asymptotically correct to the required order. With the choices
$\alpha_1 = 2/3 - \nu$, $\alpha_2 = -5/6 + 2\nu$, and $\beta = 29/72
-7\nu/6$ we obtain the equation
\[
m_t + u_x + \frac{\e_1}{2}(um_x + 2mu_x) + \frac{\e_2}{6}(1-6\nu)u_{xxx} = 0\,,
\]
where $m = u - \e_2\nu u_{xx}$.
The parameter $\nu$ in the splitting must be positive in
order to obtain correct signs of the coefficients.
To recover (\ref{CH-BV}) we reintroduce dimensional variables to
find $\alpha^2 = h^2 \nu$ and $\gamma + c_0 \alpha^2 = c_0 h^2/6$.
In these physical quantities the normalized phase speed is
\[
    \frac{\omega}{k c_0} = 1 -  \frac{h^2}{6} \frac{k^2}{1+\nu h^2 k^2} \,.
\]
The free parameter $\nu$ can now be determined by matching the
quartic term in $k$ to the exact result $\tanh(kh)/kh$, which gives $\nu =
19/60$. The original form of equation (\ref{CH-BV}) -- equation
(3) derived in \cite{CH[1993]} -- is recovered by setting $\nu = 1/3$,
which is almost the optimal value.

Equation (\ref{CH-BV}) determines the fluid velocity $u$ at a fixed height
$z_0$. To first order, the velocity $u$ and the elevation $\eta$ coincide,
but the more complicated relation (\ref{Kodama}) is required to obtain the
precise wave elevation profile.


{\it The traveling wave solution} is obtained by the usual ansatz
$u(x,t) = u(s)$, with $s = x - ct$,
after which (\ref{CH-BV}) can be integrated twice. The solution whose
velocity vanishes at spatial infinity is given by
\begin{equation}\label{TW-1}
( c+\,\gamma/\alpha^{\,2} - u)
\alpha^{\,2}(du/ds)^2
=
(c - c_0 - u)u^2
\,.
\nonumber
\end{equation}
When the limits of the radiation band coincide at zero, the peakon
traveling wave equation re-emerges. Otherwise, the traveling wave
solution can be expressed in parametric form via a Sundmann transform of
the independent variable,
$(ds/d\tau)^2=( c+\,\gamma/\alpha^{\,2} - u)\alpha^{\,2}$, as
\begin{eqnarray*}
u(\tau)  &=& u_0 {\rm sech}^2A\tau
\,,\quad
A = \sqrt{u_0} / 2
\,,\\
s(\tau)
&=& 2 \alpha \sqrt{D/u_0} \sinh^{-1}
\left( \frac{ \sinh A\tau}{\sqrt{1-u_0/D}} \right)
\\
 & &
  -2 \alpha \tanh^{-1}\left(
   1+\frac{D/u_0-1}{\tanh^2A\tau}\right)^{-1/2}
\,,
\end{eqnarray*}
where $u_0 = c - c_0  $ and $D=c_0+\gamma/\alpha^2$.
The curvature at the maximum is
$-u_0^2/2/(\gamma + c_0 \alpha^2)=-3u_0^2/(c_0h^2)$.
For $\alpha \to 0$ and $\gamma \to 0$ this smooth family approaches the
peakon solution. However, this limit cannot be attained for physical water
waves because it implies vanishing mean depth, $h \to 0$. In the physical
variables the solution reads (with rescaled parameter $\tilde\tau$)
\begin{eqnarray*}
u  &=& u_0 {\rm sech}^2 \tilde A\tilde\tau,
\quad \tilde A = \sqrt{\frac{3u_0}{2c_0h^2} }
\\
(x - ct)\tilde A
&=&
\sqrt{1+\sigma} \sinh^{-1}\left( \sqrt{1+\sigma} \sinh
\tilde A\tilde\tau \right)
\\
 & &
- \sqrt{\sigma} \tanh^{-1}\left( 1+\frac{1}{\sigma\tanh^2\tilde{A}
\tilde\tau}\right)^{-1/2} \,,
\end{eqnarray*}
where $\sigma = 6 \nu {u_0}/{c_0}$. The parameter $u_0/c_0$ plays
the same role for $u$ as
$\e_1 = a/h$ plays for $\eta$. In the KdV limit $\sigma \to 0$ the
second equation reduces to $x-ct = \tilde\tau$.
The curvature at the maximum, $-3u_0^2/(c_0h^2)$, is independent of
$\nu=\alpha^2/h^2$. Increasing $\nu$ increases the
width of the wave form without changing the curvature at its maximum.

{\it Discussion.} The water wave equation (\ref{CH-BV}) combines the
elements of linear dispersion in KdV with the nonlinear, nonlocal
dispersion of CH. This combination is IST-integrable, because it retains
the bi-Hamiltonian and isospectral properties of the KdV and CH equations.
The dispersion parameters $\alpha$, $c_0$ and $\gamma$ knit the KdV and
CH equations into a family of integrable equations and also appear together
in the eigenvalue problem (\ref{CH-BV-isospectral}).

Geometrically, equation (\ref{CH-BV}) is the  Euler-Poincar\'e equation for
geodesic motion on the Bott-Virasoro group with respect to the metric
$H_1$ as shown in \cite{Misiolek[1998]}. One may choose an arbitrary metric
for the Euler-Poincar\'e equation expressing geodesic motion on either the
diffeomorphism group or the Bott-Virasoro group, and again seek
conditions for integrability. Integrable cases of this geodesic motion for
norms other than $L_2$ and $H_1$ are not yet known, although the elastic
scattering properties of the confined pulses in some of these other cases
are discussed in \cite{Fringer&Holm[2000]}. In particular, the {\it
Galilean invariant} integrable limiting equation,
\begin{equation}\label{Distinguished-eqn}
u_{xxt} + 2u_xu_{xx} + uu_{xxx}
=
\frac{c_0}{\alpha^{2}} u_x + \frac{\gamma}{\alpha^{2}} u_{xxx}
\,,
\nonumber
\end{equation}
is obtained by taking $\alpha^{\,2}\to\infty$ in (\ref{CH-BV}) while the
ratios $c_0/\alpha^{\,2}$ and $\gamma/\alpha^{\,2}$ remain
constant. For $c_0$ and $\gamma$ absent, this equation has soliton
solutions with compact support (compactons) that were studied numerically
in \cite{Fringer&Holm[2000]}, analytically in \cite{H&Z[1994]} on the real
line, and analytically in \cite{ACHM[1994]} for the periodic case.

The dispersion parameters $c_0$ and $\gamma$ in (\ref{CH-BV}) represent
combined velocity shifts and Galilean boosts. These transformations are
{\it  deformations} of the CH equation -- they are not its symmetries.
While these dispersive deformations preserve the isospectral
property of the CH equation, they alter its isospectral content in two
ways. First, they add to the discrete CH peakon spectrum a band of
continuous spectrum near the origin. This isospectral band corresponds to
linear radiation and spans the allowed range of linear phase velocity.
Second, either $c_0$ or $\gamma$ nonvanishing breaks the
reflection-reversal symmetry of the CH equation for $c_0=0$ that allows
coexistence of its generalized-function peakon and antipeakon soliton
solutions traveling in opposite directions. Thus, adding linear dispersion
breaks this reflection-reversal symmetry and removes the antipeakon
solutions, while it also rounds the peaks of the peakon solutions, thereby
regularizing them into ordinary solitons, plus linear radiation.

This letter reconfirmed that (\ref{CH-BV}) is a genuine shallow water
equation by rederiving it via asymptotic expansion of the Euler
equations. This derivation used a nonlocal transformation due to
Kodama \cite{Kodama[1985]} as well as a splitting of the time derivative
at the highest order that introduced an optimization parameter,
$\nu=\alpha^2/h^2$. We showed that the equation originally discovered in
\cite{CH[1993]} via asymptotic expansion of the Euler fluid Hamiltonian
is contained in the family of equations derived here for the case
$\nu=1/3$. Moreover, the value $\nu=1/3$ obtained in the Hamiltonian
derivation in \cite{CH[1993]} differs only very little from the optimal
value $\nu=19/60$ yielding the most accurate linear
dispersion relation among those in the family derived here. We also showed
that varying $\nu$ alters the width of the traveling wave profile for
equation (\ref{CH-BV}) without changing the curvature at its maximum.

{\bf Acknowledgments.}  
We are grateful to R. Camassa, O. Fringer, I. Gabitov, R. Grimshaw,
B. McCloud, J. Montaldi, P. J. Olver, T. Ratiu, M. Roberts, P. Swart
and A. Zenchuk for their constructive comments and encouragement. This
work began at the MASEES 2000 Summer School in Peyresq, France.  DDH
was supported by DOE, under contract W-7405-ENG-36.


\end{multicols}

\end{document}